\documentclass[a4paper,12pt]{article}
\usepackage{amsthm}
\usepackage{amssymb}
\usepackage{amsmath}
\usepackage{amsfonts}
\usepackage{color}
\usepackage{graphicx}
\usepackage[square, comma, sort&compress, numbers]{natbib}

\newtheorem{remark}{Remark}[section]

\newtheorem{lemma}{Lemma}[section]
\newtheorem{theorem}{Theorem}[section]

\newtheorem{corollary}{Corollary}[section]

\def\b1{\mbox{\boldmath $1$}}

\newenvironment{demo*}{\vspace{3mm}\noindent{\bf Proof.}}{\hfill $\Box$ \vspace{3mm}}

\begin{document}
\title{\bf \Large {Remarks on  equality of two distributions under some partial orders}}
{\color{red}{\author{%\normalsize{Ying Zhang}\\{\normalsize\it  School of Mathematical Sciences, Qufu Normal University}\\\noindent{\normalsize\it Shandong 273165, China}\\e-mail:  zhangying0513@gmail.com\\\and
\normalsize{Chuancun Yin}\\
%\thanks{Corresponding author.}\\
{\normalsize\it  School of Statistics,  Qufu Normal University}\\
\noindent{\normalsize\it Shandong 273165, China}\\
e-mail:  ccyin@mail.qfnu.edu.cn}}}%\\
%\and
%\normalsize{Ming Zhou}\\{\normalsize\it  China Institute for Actuarial Science, Central University of Finance and Economics}\\
%\noindent{\normalsize\it Beijing 100081, China}\\
%e-mail:  mzhou.act@gmail.com
\maketitle
\vskip0.01cm
%\newpage
\noindent{ {\bf Abstract}}  { { In this note  we establish some   appropriate conditions  for stochastic equality of two  random variables/vectors
which are ordered with respect to convex  ordering or with respect to supermodular ordering. Multivariate extensions of this result are also considered. }}

\medskip


%\newpage

%\noindent{\bf 1.~~Introduction}
\numberwithin{equation}{section}
\section{INTRODUCTION}\label{intro}

Let $X$ and $Y$ be two  random variables with distribution functions $F_X$ and $F_Y$ respectively. Let $\overline{F}_X$ and $\overline{F}_Y$  denote
the corresponding survival functions. $X$ is said to precede $Y$ in the stop-loss order sense, notation $X\le_{sl} Y$, if and only if $E[(X-d)_+]\le E[(Y-d)_+], -\infty<d<\infty$; $X$ is said to precede $Y$ in the convex order sense, notation $X\le_{cx} Y$, if and only if  $X\le_{sl} Y$ and in addition $E[X]=E[Y]$.
Equivalently, $X\le_{cx}Y$ if and only if $Ef(X)\le Ef(Y)$ for every convex function $f$, provided that expectations $Ef(X)$ and  $Ef(Y)$ exist.
The stop-loss order can be characterized in terms of ordered TVaR's (see e.g. Dhaene et al. (2006)):
 $X\le_{sl} Y \Leftrightarrow TVaR_p[X]\le  TVaR_p[Y]$ for all $p\in (0,1)$, where
 $TVaR_p[X]=\frac{1}{1-p}\int_p^1 F_X^{-1}(q)dq$ is the tail value-at -risk at level $p$, and $F_X^{-1}(q)=\inf\{x\in \Bbb{R}|F_X(x)\ge q\}$ with $\inf\emptyset=+\infty$, by convention. A random vector $Y=(Y_1,\cdots, Y_n)$  with marginal distributions
$F_{Y_i}, i = 1, 2, \cdots, n$, is called commonotonic
 if
 $$Y\stackrel{d}{=}(F_{Y_1}^{-1}(U), F_{Y_2}^{-1}(U),\cdots, F_{Y_n}^{-1}(U)),$$
 where $\stackrel{d}{=}$ stands for ``equality in distribution" and $U$ is a random variable that is uniformly distributed over the unit interval $(0,1)$.
 Consider a random vector $(Y_1, \cdots, Y_n)$  and its comonotonic counterpart $(Y_1^c, \cdots, Y_n^c)$.
 The sum of the components is denoted by $S$ and $S^c$ respectively. A nice result of Kaas et al. (2002) says that $S\le_{cx} S^c$,  and the converse remains valid by Theorem 4 in Cheung (2010);  see Mao and Hu (2011) for a new proof.
 For more details about  comonotonicity, stochastic orders and their applications, we refer the reader, e.g., to  Joe (1997), Shaked and Shanthikumar (2007) and Denuit et al. (2005).

 Cheung (2010) proved the following theorems giving sufficient conditions for stochastic equality of  two
random variables when these are known to be stochastically ordered.

\begin{theorem} (Cheung (2010), Theorem 6) \ Let $Y_1$ and $Y_2$ be two integrable random variables and $u$
be any  real-valued strictly convex   function or strictly concave function which is twice continuously differentiable.
Then
$$Y_1\le_{cx}Y_2  \; {\rm and}\;  E[u(Y_1)]= E[u(Y_2)]\Rightarrow Y_1\stackrel{d}{=}Y_2.$$
In particular,
 $$E[u(S)]= E[u(S^c)]\Leftrightarrow S\stackrel{d}{=}S^c.$$
\end{theorem}

\begin{theorem} (Cheung (2010), Theorem 7) \ Let $Y_1$ and $Y_2$ be two integrable random variables, and $g$
be a strictly concave continuously differentiable  distortion  function with $g'(0)<\infty$.
Then
$$Y_1\le_{cx}Y_2  \; {\rm and}\;  \rho_g[Y_1]= \rho_g[Y_2]\Rightarrow Y_1\stackrel{d}{=}Y_2.$$
In particular,
 $$ \rho_g[S]= \rho_g[S^c]\Leftrightarrow S\stackrel{d}{=}S^c.$$
\end{theorem}
Cheung et al. (2015, Theorem 7) obtained the same result as in Theorem 1.1 under the following weaker conditions on $u$: $u$ is  a strictly concave (or strictly convex) function with absolutely continuous derivative $u'$. Cheung et al. (2015, Theorem 8) obtained the same result as in Theorem 1.2 under the following more general conditions on  the distortion function $g$: $g$ is continuously differentiable and  strictly concave (or strictly convex).
We remark that there is  a very minor gap in the proof to Theorems 7  and 8 in  Cheung et al. (2015). Our aim in this paper is to fill this gap and obtain more general sufficient conditions for stochastic equality of two  random variables/vectors which
are ordered with respect to the partial orders.

The rest of the paper is organized as follows. We review some basic definitions and notations  such as  convex and concave functions in Section 2. In Section 3  we characterize comonotonicity by distortion risk measures, and in Section 4 we  characterize comonotonicity through expected utility.
Finally, in Section 5 the multivariate extensions are considered.

\numberwithin{equation}{section}
\section{Some results for convex and concave functions}\label{intro}

Throughout the paper, we will use the notion $I$ to denote a nondegenerate interval of the real line.
In this section, we present several concepts and results that will be used throughout the
paper.

{\bf Definition 2.1}\;  A function $f : I \rightarrow \Bbb{R}$ is called convex if
\begin{equation}
f((1-\lambda)x+\lambda y)\le (1-\lambda)f(x)+\lambda f(y)
\end{equation}
for all points $x$ and $y$ in $I$ and all $\lambda \in [0, 1]$.
It is called strictly convex if
the inequality (2.1) holds strictly whenever $x$ and $y$ are distinct points and $\lambda \in [0, 1]$.
 If -$f$ is convex (respectively, strictly convex) then we say that $f$ is
concave (respectively, strictly concave).

Here are several elementary examples of  convex functions of one variable:

$\bullet$\; functions convex on the whole axis:
$x^{2r}, r {\rm \; being \; positive\;  integer;}\; e^{tx}, t\neq 0; (x-a)^2, a\in\Bbb{R}.$

$\bullet$\; functions convex on the nonnegative ray: $x^r, r\ge 1; -x^r, 0\le r\le 1; x\ln x.$

$\bullet$\; functions convex on the positive ray: $x^{-r}, r>0; -\ln x.$

The following lemma is the   result on the smoothness of convex
functions, which can be found in Niculescu and Persson (2006, P. 21).
\begin{lemma}
Let $f : I\rightarrow\Bbb{R}$  be a convex function. Then
$f$ is continuous on the interior int($I$) of $I$ and has finite left and right derivatives
at each point of int($I$). Moreover,$x < y$ in int($I$) implies
$$f_{-}'(x)\le f_{+}'(x)\le f_{-}'(y)\le f_{+}'(y)$$
Particularly, both $f_{-}'$ and $f_{+}'$ are nondecreasing on int($I$).
\end{lemma}
A convex function $f$ defined on some open interval $I$ is continuous on $I$ and Lipschitz continuous on any closed subinterval. $f$ admits left and right derivatives, and these are monotonically non-decreasing. As a consequence, $f$ is differentiable at all but at most countably many points. If $I$ is closed, then $f$ may fail to be continuous at the endpoints of $I$. For example,
the function $f$ with domain [0,1] defined by $f(0) = f(1) = 1, f(x) = 0$ for $0<x<1$ is convex; it is continuous on the open interval $(0,1)$, but not continuous at 0 and 1.
\begin{lemma}
(The second derivative test)\; Suppose that $f : I \rightarrow \Bbb{R}$ is a twice differentiable function. Then:\\
(i) $f$ is convex if and only if $f''\ge 0$;\\
(ii) $f$ is strictly convex if and only if $f''\ge 0$  and the set of points where $f''$
vanishes does not include intervals of positive length.
\end{lemma}
A proof of this result can be found e.g., in  Niculescu and Persson (2006).
\begin{remark} An important result due to A. D. Alexandrov asserts that all convex functions
are almost everywhere twice differentiable. See Theorem 3.11.2. in  Niculescu
and Persson (2006). Riesz-Nagy gave an example of real-valued function $\phi$ on $[0,1]$ such that $\phi(0)=0, \phi(1)=1$, $\phi$ is continuous and  strictly increasing, and
 $\phi' = 0$
almost everywhere. See Hewitt and Stromberg (1965, Example 18.8, p. 278). Thus the  function $u(x)=\int_0^x \phi(t)dt$
 is strictly convex though $u'' = 0$ almost everywhere; see  Niculescu
and Persson (2006, P. 37).
\end{remark}

 \vskip 0.2cm
 \section{Convex order, expected utility and comonotonicity}
\setcounter{equation}{0}

{\bf Definition 2.1}\ Let two measures $P$ and $Q$ be defined on the same space. $Q$ is called absolutely continuous with respect to $P$, written as $Q\ll P$, if $Q(A)=0$ whenever $P(A)=0$ for any measurable set $A$. $P$ and $Q$ are called equivalent if $Q\ll P$ and $P\ll Q$.

\begin{theorem} \ Let $Y_1$ and $Y_2$ be two integrable random variables on interval $I$, and $u : I\rightarrow \Bbb{R}$
be any   convex   function. Assume that $\lambda \ll \gamma$, where $\lambda$ is the Lebesgue
measure on $\Bbb{R}$ and $\gamma$ is the positive Radon measure defined by
$$\gamma(x,y]=u'_{+}(y)-u'_{+}(x)\; {\rm for\; any}\; x<y,$$
where $u'_{+}$ is the  right-hand derivative of $u$.
Then
$$Y_1\le_{cx}Y_2  \; {\rm and}\;  E[u(Y_1)]= E[u(Y_2)]\Rightarrow Y_1\stackrel{d}{=}Y_2.$$
In particular,
 $$E[u(S)]= E[u(S^c)]\Leftrightarrow S\stackrel{d}{=}S^c.$$
\end{theorem}
By switching from $u$ to $-u$, yields that
\begin{corollary}
\ Let $Y_1$ and $Y_2$ be two integrable random variables on interval $I$, and $u : I\rightarrow \Bbb{R}$
be any   concave   function. Assume that $\lambda \ll \gamma$, where $\lambda$ is the Lebesgue
measure on $\Bbb{R}$ and $\gamma$ is the positive Radon measure defined by
$$\gamma(x,y]=u'_{+}(x)-u'_{+}(y)\; {\rm for\; any}\; x<y,$$
where $u'_{+}$ is the  right-hand derivative of $u$.
Then
$$Y_1\le_{cx}Y_2  \; {\rm and}\;  E[u(Y_1)]= E[u(Y_2)]\Rightarrow Y_1\stackrel{d}{=}Y_2.$$
In particular,
 $$E[u(S)]= E[u(S^c)]\Leftrightarrow S\stackrel{d}{=}S^c.$$
\end{corollary}
\begin{remark}
%Cheung (2010) obtained the above results under conditions that $u$ is a any real-valued strictly convex or strictly concave function which is twice continuously differentiable instead of our condition $\lambda \ll \gamma$ (see Cheung (2010, Theorem 6)).
 If $u$ is  convex   and  $u''>0$  almost everywhere or  if $u$ is  concave   and  $u''<0$  almost everywhere, or, more generally,  $u$ is a any real-valued  strictly convex or strictly concave function,  then $\gamma$ is equivalent to $\lambda$.  Thus Theorem 3.1 and Corollary 3.1 is generalization    of Theorem 1.1.
\end{remark}
\begin{remark} %Theorem 7 in  Cheung et al. (2015) generalizes Cheung's result  for a wilder class of strictly concave (or strictly convex) function $u$ with absolutely continuous derivative $u'$. This result, as pointed out in Cheung et al. (2015), also follows from Theorem 3A43 in Shanthikumar and Shaked (2007) which states that: suppose that $X\le_{cx}Y$ and that for some strictly convex function $\phi$ we have that $E\phi(X)=E\phi(Y)$, provided the expectations exist. Then $X\stackrel{d}{=}Y$.
We remark that
the proof to Theorem 7 in  Cheung et al. (2015)   has a gap if there is no further restrictions on $u$ (for example,  $u''>0$ a.e. on $I$). In fact, the function $u$ in Remark 2.1 is an example of a strictly convex but  $u'' = 0$ almost everywhere.
\end{remark}

The proof of Theorem 3.1 requires    the following lemma, which  can be found in F\"{o}llmer and Schied (2004), see also Cheung (2010).

\begin{lemma}
 Suppose that $u$ is an increasing convex function  with right-hand
derivative $u'_{+}$. There is a positive Radon measure $\gamma$
 on $\Bbb{R}$  such that
 $$\gamma(x,y]=u'_+(y)-u'_+(x)\; {\rm for\; any}\; x<y, $$
 and
 \begin{eqnarray*}
 u(x)&=&u(0)+u'(0)x+\int_{(0,\infty)}(x-t)_+ \gamma(dt)\\
 &&+\int_{(-\infty,0]}(t-x)_+ \gamma(dt),\; x\in\Bbb{R}.
 \end{eqnarray*}
\end{lemma}
{\bf Proof of Theorem 3.1.}\ We prove the theorem for  the case where $u$ is increasing convex function  only, the rest cases can be handled in a similar way as the proof to Theorem 6 in Cheung (2010). Notice that the convex order relation  $Y_1\le_{cx}Y_2$ implies that $E[Y_1]=E[Y_2]$. As in the step 1 of the  proof to Theorem 6  in Cheung (2010), the condition $E[u(Y_1)]= E[u(Y_2)]$ imply that
\begin{eqnarray*}
\int_{(0,\infty)}\left\{E(Y_2-t)_{+}-E(Y_1-t)_{+}\right\}\gamma(dt)+\int_{(-\infty,0]}\left\{E(t-Y_2)_{+}-E(t-Y_1)_{+}\right\}\gamma(dt)=0.
 \end{eqnarray*}
 Since $Y_1\le_{cx} Y_2$, we  have $E(Y_2-t)_{+}-E(Y_1-t)_{+}\ge 0$ and $E(t-Y_2)_{+}-E(t-Y_1)_{+}\ge 0$ for all $t$. It follows that
 $E(Y_2-t)_{+}=E(Y_1-t)_{+}$ for $\gamma$-almost all $t>0$ and $E(t-Y_2)_{+}=E(t-Y_1)_{+}$ for $\gamma$-almost all $t\le 0$, and hence
 $E(Y_2-t)_{+}=E(Y_1-t)_{+}$ for $\lambda$-almost all $t>0$ and $E(t-Y_2)_{+}=E(t-Y_1)_{+}$ for $\lambda$-almost all $t\le 0$
 since $\lambda \ll \gamma$. As the functions $E(Y_i-t)_{+}$ and $E(t-Y_i)_{+}$ are continuous functions of $t$, we conclude that  $Y_1$ and $Y_2$ have the same distribution.

 \vskip 0.2cm
 \section{Convex order, distorted expectations and comonotonicity}
\setcounter{equation}{0}

A distortion function is a non-decreasing function $g: [0,1]\rightarrow [0,1]$ such that $g(0) = 0$
and $g(1) = 1$.
The distorted expectation of the random variable $X$ associated with distortion
function $g$, notation $\rho_{g}[X]$,  is defined as
$$\rho_{g}[X]=\int_0^{+\infty}g(\bar{F}_X(x))dx+\int_{-\infty}^0 [g(\bar{F}_X(x))-1]dx,$$
provided at least one of the to integrals above is finite. If $X$ a non-negative  random variable, then $\rho_{g}$ reduces to
$$\rho_{g}[X]=\int_0^{+\infty}g(\bar{F}_X(x))dx.$$
 In view of Dhaene et al. (2012, Theorems 4 and 6) we know  that, when the distortion function $g$ is right continuous
on $[0,1)$, then  $\rho_{g}[X]$ may be rewritten as
$$\rho_{g}[X]=\int_{[0,1]}VaR^+_{1-q}[X]dg(q),$$
where
$VaR^+p[X]=\sup\{x|F_X(x)\le p\}$, and when the distortion function $g$ is left continuous
on $(0,1]$, then  $\rho_{g}[X]$ may be rewritten as
$$\rho_{g}[X]=\int_{[0,1]}VaR_{1-q}[X]dg(q)=\int_{[0,1]}VaR_{q}[X]d{\bar g}(q),$$
where
$VaR_p[X]=\inf\{x|F_X(x)\ge p\}$ and ${\bar g}(q):=1-g(1-q)$ is the dual distortion of $g$. Obviously, ${\bar {\bar g}}=g$,
$g$ is left continuous if and only if ${\bar g}$ is right continuous; $g$ is concave  if and only if ${\bar g}$ is convex.

\begin{theorem} \ Let $Y_1$ and $Y_2$ be two integrable random variables, and $g$
be a  concave   distortion  function. Assume that $\lambda \ll \nu$, where $\lambda$ is the Lebesgue
measure on $\Bbb{R}$ and $\nu$ is the  Radon measure defined by
  $\nu([0,q])=g'_+(1-q)$.
Then we have that
$$Y_1\le_{cx}Y_2  \; {\rm and}\;  \rho_g[Y_1]= \rho_g[Y_2]\Rightarrow Y_1\stackrel{d}{=}Y_2.$$
In particular,
 $$ \rho_g[S]= \rho_g[S^c]\Leftrightarrow S\stackrel{d}{=}S^c.$$
\end{theorem}

{\bf Proof}\ The distortion measure with concave distortion function $g$ can be expressed  by the  weighted  TVaR. In fact, note that $\phi(q)=g'_+(1-q)$ is monotone increasing, so $\nu([0,q])=\phi(q)$ is positive measure.  We have
\begin{eqnarray}
\rho_{g}[X]&=&-\int_0^1  VaR_w[X] dg(1-w)\nonumber\\
&=&\int_0^1  VaR_w[X] g'_+(1-w)dw\nonumber\\
&=&\int_0^1  VaR_w[X] \phi(w)dw\nonumber\\
&=&\nu([0,1])EX+\int_0^1 TVaR_w[X](1-w)d\nu(w)\nonumber\\
&=&\nu([0,1])EX+\int_0^1 TVaR_w[X]d\mu(w),
\end{eqnarray}
where
$$d\mu(w)=(1-w)d\nu(w).$$ It can be shown that $\mu$ is a probability measure. In fact,
\begin{eqnarray*}
\int_0^1 d\mu(w)&=&\int_0^1 \nu([0,w])dw\\
&=&\int_0^1 \phi(w)dw=\int_0^1 g'_+(w)dw=1.
\end{eqnarray*}
The convex order $Y_1\le_{cx}Y_2$ implies that
$EY_1=EY_2$ and
$TVaR_p[Y_1]\le TVaR_p[Y_2]$, for all $p\in (0,1)$.
As in Cheung et al. (2015) we have
\begin{eqnarray*}
0&=&\rho_g[Y_1]-\rho_g[Y_1]\\
&=&\int_0^1 TVaR_w[Y_1](1-w)d\nu(w)-\int_0^1 TVaR_w[Y_2](1-w)d\nu(w).
\end{eqnarray*}
We conclude that
$TVaR_p[Y_1]=TVaR_p[Y_2]$,  for $\nu$-almost all   $p\in (0,1)$, and hence $TVaR_p[Y_1]=TVaR_p[Y_2]$,  for $\lambda$-almost all   $p\in (0,1)$ since $\lambda \ll \nu$.
Furthermore, as the function $TVaR_p[Y_1]-TVaR_p[Y_2]$ is a continuous function of $p$, we have  $TVaR_p[Y_1]=TVaR_p[Y_2]$,  for all   $p\in (0,1)$,
 which is equivalent with
$E(Y_2-t)_{+}=E(Y_1-t)_{+}$ for  all $t\in \Bbb{R}$. Thus $Y_1\stackrel{d}{=}Y_2.$

\begin{corollary} \ Let $Y_1$ and $Y_2$ be two integrable random variables, and $g$
be a strictly convex   distortion  function. Assume that $\lambda \ll \nu$, where $\lambda$ is the Lebesgue
measure on $\Bbb{R}$ and $\nu$ is the  Radon measure defined by
  $\nu([0,q])=-g'_+(1-q)$.
Then we have that
$$Y_1\le_{cx}Y_2  \; {\rm and}\;  \rho_g[Y_1]= \rho_g[Y_2]\Rightarrow Y_1\stackrel{d}{=}Y_2.$$
In particular,
 $$ \rho_g[S]= \rho_g[S^c]\Leftrightarrow S\stackrel{d}{=}S^c.$$
\end{corollary}

\begin{remark}
%Cheung (2010) obtained the above results under conditions that $g$ is a   strictly concave, continuously differentiable distortion function, and $g'(0)<\infty$  (see Cheung (2010, Theorem 7)).
If   $g$ is a any real-valued  strictly convex or strictly concave function,  then $\nu$ is equivalent to $\lambda$.  Thus Theorem 4.1 is generalization    of Theorem 1.2.
\end{remark}
\begin{remark} Theorem 8 in  Cheung et al. (2015) obtained the above results under conditions that
 $g$ is a   strictly concave (or strictly convex) distortion function with absolutely continuous derivative $g'$.
   We remark that, as in Remark 3.2,
the proof to Theorem 8 in  Cheung et al. (2015)   has a minor gap if there is no further restrictions on $g$ (for example,  $g''>0$ a.e. on $[0,1]$).
 In fact,  the  function $g(x)=\frac{\int_0^x \phi(t)dt} {\int_0^1 \phi(t)dt}$
 is strictly increasing  distortion function, but $u'' = 0$ almost everywhere, where  the function $\phi$ is defined  in Remark 2.1.
\end{remark}

\begin{remark} As remarked  in Cheung (2010), the condition $Y_1\le_{cx}Y_2$ in Theorem 4.1 can be slightly relaxed to $Y_1\le_{sl}Y_2$.
In fact,  as is well known $Y_1\le_{sl}Y_2$ implies that $EY_1\le EY_2$. Moreover,  $Y_1\le_{sl}Y_2 \Leftrightarrow TVaR_p[Y_1]\le TVaR_p[Y_2]$ for all $p\in (0,1)$ (see, e.g. Dhaene et al. (2006), Theorem 3.2). By using (4.1) we have
\begin{eqnarray*}
0&=&\rho_g[Y_1]-\rho_g[Y_1]\\
&=& \nu([0,1])(EY_1-EY_2)+\int_0^1 \left(TVaR_w[Y_1]-TVaR_w[Y_2]\right)d\mu(w),
\end{eqnarray*}
which implies $EY_1=EY_2$ and $\int_0^1 \left(TVaR_w[Y_1]-TVaR_w[Y_2]\right)d\mu(w)=0$. Thus   $Y_1\stackrel{d}{=}Y_2.$
\end{remark}

 \vskip 0.2cm
 \section{Multivariate extensions}
\setcounter{equation}{0}

As in Cheung et al. (2015) we use the notions $\underline{X}$ and $\underline{Y}$ to denote the $n$-vectors $(X_1, X_2,\cdots, X_n)$ and
$(Y_1, Y_2,\cdots, Y_n)$, respectively. The sums of their components are denoted by $S_X$ and $S_Y$, respectively:
$$S_X=X_1+\cdots+X_n,\;\;{\rm and}\;\; S_Y=Y_1+\cdots+Y_n.$$
{\bf Definition 5.1}\; A function $f:\Bbb{R}^n\rightarrow \Bbb{R}$ is said to be supermodular if for any $\underline{X}, \underline{Y}\in\Bbb{R}^n$ it satisfies
$$f(\underline{X})+f(\underline{Y})\le f(\underline{X}\wedge \underline{Y})+f(\underline{X}\vee\underline{Y}),$$
where the operators $\vee$ and $\wedge$ denote coordinatewise minimum and maximum, respectively. $\underline{X}$ is said to be smaller in the supermodular order that $\underline{Y}$, notation $\underline{X}\le_{SM}\underline{Y}$, if $Ef(\underline{X})\le Ef(\underline{Y})$ holds for all supermodular functions $f:\Bbb{R}^n\rightarrow \Bbb{R}$ for which the expectations exist.

Parallel to the Theorems 13 and 14 in Cheung et al. (2015), we have the following two theorems under weaker conditions on $u$ and $g$.
\begin{theorem} Consider the $n$-vectors $\underline{X}$ and $\underline{Y}$ with respective sums $S_X$ and $S_Y$ which are assumed to have finite expectations. Furthermore, consider the interval $I$ with $P(S_Y\in I)=1,$
and $u : I\rightarrow \Bbb{R}$
be any   concave   function. Assume that $\lambda \ll \gamma$, where $\lambda$ is the Lebesgue
measure on $\Bbb{R}$ and $\gamma$ is the positive Radon measure defined by
$$\gamma(x,y]=u'_{+}(x)-u'_{+}(y)\; {\rm for\; any}\; x<y,$$
where $u'_{+}$ is the  right-hand derivative of $u$. Finally, suppose that either
$$E[\max(S_Y, 0)]^{n-1}<\infty\;\; {\rm or}\;\; E[-\min(S_Y, 0)]^{n-1}<\infty.$$
Then we have that
$$\underline{X}\le_{SM}\underline{Y}\;{\rm and}\;  E[u(S_X)]= E[u(S_Y)]\Rightarrow \underline{X} \stackrel{d}{=}\underline{Y}.$$
\end{theorem}
\begin{theorem} Consider the $n$-vectors $\underline{X}$ and $\underline{Y}$ with respective sums $S_X$ and $S_Y$ which are assumed to have finite expectations. Furthermore, let $g$ be a  concave  distortion function.  Assume that $\lambda \ll \nu$, where $\lambda$ is the Lebesgue
measure on $\Bbb{R}$ and $\nu$ is the  Radon measure defined by
  $\nu([0,q])=g'(1-q)$.
Finally, suppose that either
$$E[\max(S_Y, 0)]^{n-1}<\infty\; \; {\rm or}\;\; E[-\min(S_Y, 0)]^{n-1}<\infty.$$
Then we have that
$$\underline{X}\le_{SM}\underline{Y}\;{\rm and}\;  \rho_g[S_X]=\rho_g[S_Y]\Rightarrow \underline{X} \stackrel{d}{=}\underline{Y}.$$
\end{theorem}
\begin{remark} By switching from $u$ to $-u$ in Theorems 5.1 and 5.2 we can obtain the  versions for  convex  functions $u$ and $g$.
\end{remark}

\noindent{\bf Acknowledgements.} \ % The authors are grateful to the anonymous referee's careful reading and detailed helpful comments and constructive suggestions, which have led to a significant improvement of the paper.
The research   was supported by the National
Natural Science Foundation of China (No. 11171179),  the Research
Fund for the Doctoral Program of Higher Education of China (No. 20133705110002) and the Program for  Scientific Research Innovation Team in Colleges and Universities of Shandong Province.

\end{document}